%% file: skeleton.tex
\documentclass{PoS}
\usepackage{ifthen}

\newboolean{pdflatex}
\setboolean{pdflatex}{true} % False for eps figures 

\newboolean{articletitles}
\setboolean{articletitles}{true} % False removes titles in references

\newboolean{uprightparticles}
\setboolean{uprightparticles}{false} %True for upright particle symbols

\newboolean{inbibliography}
\setboolean{inbibliography}{false} %True once you enter the bibliography

\newboolean{uselhcbcpconvention}
\setboolean{uselhcbcpconvention}{true}

\usepackage{microtype}
\usepackage{lineno}  % for line numbering during review
\usepackage{xspace} % To avoid problems with missing or double spaces after
\usepackage{caption} %these three command get the figure and table captions automatically small

\usepackage{graphicx}  % to include figures (can also use other packages)
\usepackage{color}
\usepackage{colortbl}

\usepackage{amsmath} % Adds a large collection of math symbols
\usepackage{amssymb}
\usepackage{amsfonts}
\usepackage{upgreek} % Adds in support for greek letters in roman typeset

\input{lhcb-symbols-def} % Add in the predefined LHCb symbols

\input{ourdefs}

\title{Summary of CKM 2016 Working Group 4 : mixing and mixing-related $CP$ violation in the $B$ system}

\ShortTitle{Summary of CKM 2016 Working Group 4}

\author{\speaker{Alessandro Gaz}\\
        Kobayashi-Maskawa Institute, Nagoya University, Nagoya, Japan\\
        E-mail: \email{gaz@hepl.phys.nagoya-u.ac.jp}}
\author{
        Vladimir V. Gligorov\\%\thanks{I would like to thank my mum, dad, and trade union representative for their support. Couldn't have done it without you!}\\
        LPNHE, Universit\'{e} Pierre et Marie Curie, Universit\'{e} Paris Diderot, CNRS/IN2P3, Paris, France\\
        E-mail: \email{vgligoro@lpnhe.in2p3.fr}
        }
\author{
        Dean Robinson\\%\thanks{Show me an ambulance dashing off into the night, and I shall show you the theorists bravely in pursuit.}\\
        University of Cincinnati, Cincinnati, Ohio, USA\\
        E-mail: \email{dean.robinson@uc.edu}\\
        }

\abstract{We summarize the new results on $B$ meson mixing and mixing-related $CP$ violation presented at CKM2016.
We place these results in the context of both previous experimental measurements and the most recent theoretical developments
in the field, and discuss prospects for ongoing and future mixing and mixing-related $CP$ violation measurements.}

\FullConference{9th International Workshop on the CKM Unitarity Triangle\\
		28  November - 3 December 2016\\
		Tata Institute for Fundamental Research (TIFR), Mumbai, India}

\begin{document}

\input{introduction}

%\input{phis}
%\input{deltag}
%\input{deltam}
% ale: I think it makes more sense to have phi_s and deltaGamma_s in one section:
% vava : agree on all counts
\input{phis-deltags}
% and one section for Deltam_d,s and DeltaGamma_d:
\input{deltam-deltagd}

% (this reflects better the structure of the talks IMHO)
\input{photpol}
\input{alpha}
\input{beta}
\input{gamma}
\input{theory}

\input{conclusion}

\bibliographystyle{jhep}
\bibliography{skeleton}

\end{document}

%% file: lhcb-symbols-def.tex
\usepackage{xspace} 
\usepackage{upgreek}

\def\MagUp {\mbox{\em Mag\kern -0.05em Up}\xspace}

\ifthenelse{\boolean{uprightparticles}}%
{

 \def\Ppsi        {\ensuremath{\uppsi}\xspace}

 \def\PDelta      {\ensuremath{\Delta}\xspace}                 
 \def\PXi      {\ensuremath{\Xi}\xspace}                 
 \def\PLambda      {\ensuremath{\Lambda}\xspace}                 
 \def\PSigma      {\ensuremath{\Sigma}\xspace}                 
 \def\POmega      {\ensuremath{\Omega}\xspace}                 
 \def\PUpsilon      {\ensuremath{\Upsilon}\xspace}                 
 
 \def\PB      {\ensuremath{\mathrm{B}}\xspace}                 
                  
 \def\PD      {\ensuremath{\mathrm{D}}\xspace}

 \def\PJ      {\ensuremath{\mathrm{J}}\xspace}                 
 \def\PK      {\ensuremath{\mathrm{K}}\xspace}

 \def\Pi      {\ensuremath{\mathrm{i}}\xspace}

 \def\Ps      {\ensuremath{\mathrm{s}}\xspace}

}
{

 \def\Ppsi        {\ensuremath{\psi}\xspace}                 
                  
 \mathchardef\PDelta="7101
 \mathchardef\PXi="7104
 \mathchardef\PLambda="7103
 \mathchardef\PSigma="7106
 \mathchardef\POmega="710A
 \mathchardef\PUpsilon="7107
                  
 \def\PB      {\ensuremath{B}\xspace}                 
                  
 \def\PD      {\ensuremath{D}\xspace}

 \def\PJ      {\ensuremath{J}\xspace}                 
 \def\PK      {\ensuremath{K}\xspace}

 \def\Pi      {\ensuremath{i}\xspace}

 \def\Ps      {\ensuremath{s}\xspace}

}

\makeatletter
\ifcase \@ptsize \relax% 10pt
  \newcommand{\miniscule}{\@setfontsize\miniscule{4}{5}}% \tiny: 5/6
\or% 11pt
  \newcommand{\miniscule}{\@setfontsize\miniscule{5}{6}}% \tiny: 6/7
\or% 12pt
  \newcommand{\miniscule}{\@setfontsize\miniscule{5}{6}}% \tiny: 6/7
\fi
\makeatother

\DeclareRobustCommand{\optbar}[1]{\shortstack{{\miniscule (\rule[.5ex]{1.25em}{.18mm})}
  \\ [-.7ex] $#1$}}

\def\squark    {{\ensuremath{\Ps}}\xspace}

\def\kaon    {{\ensuremath{\PK}}\xspace}
  \def\Kbar    {{\kern 0.2em\overline{\kern -0.2em \PK}{}}\xspace}

\def\KorKbar    {\kern 0.18em\optbar{\kern -0.18em K}{}\xspace}

\def\Kp      {{\ensuremath{\kaon^+}}\xspace}

  \def\Dbar    {{\kern 0.2em\overline{\kern -0.2em \PD}{}}\xspace}
\def\D       {{\ensuremath{\PD}}\xspace}

\def\DorDbar    {\kern 0.18em\optbar{\kern -0.18em D}{}\xspace}

\def\Ds      {{\ensuremath{\D^+_\squark}}\xspace}

\def\B       {{\ensuremath{\PB}}\xspace}
\def\Bbar    {{\ensuremath{\kern 0.18em\overline{\kern -0.18em \PB}{}}}\xspace}

\def\BorBbar    {\kern 0.18em\optbar{\kern -0.18em B}{}\xspace}

\def\Bs      {{\ensuremath{\B^0_\squark}}\xspace}

\def\jpsi     {{\ensuremath{{\PJ\mskip -3mu/\mskip -2mu\Ppsi\mskip 2mu}}}\xspace}

  \def\Y#1S{\ensuremath{\PUpsilon{(#1S)}}\xspace}% no space before {...}!

\def\Lbar        {{\ensuremath{\kern 0.1em\overline{\kern -0.1em\PLambda}}}\xspace}
\def\LorLbar    {\kern 0.18em\optbar{\kern -0.18em \PLambda}{}\xspace}

\def\to                 {\ensuremath{\rightarrow}\xspace}

\def\CP                {{\ensuremath{C\!P}}\xspace}

\newcommand{\phis}{{\ensuremath{\phi_{\squark}}}\xspace}

\def\AT#1     {\ensuremath{A_{\mathrm{T}}^{#1}}\xspace}           % 2

\def\C#1      {\ensuremath{\mathcal{C}_{#1}}\xspace}                       % 9
\def\Cp#1     {\ensuremath{\mathcal{C}_{#1}^{'}}\xspace}                    % 7
\def\Ceff#1   {\ensuremath{\mathcal{C}_{#1}^{\mathrm{(eff)}}}\xspace}        % 9  
\def\Cpeff#1  {\ensuremath{\mathcal{C}_{#1}^{'\mathrm{(eff)}}}\xspace}       % 7
\def\Ope#1    {\ensuremath{\mathcal{O}_{#1}}\xspace}                       % 2
\def\Opep#1   {\ensuremath{\mathcal{O}_{#1}^{'}}\xspace}                    % 7

\newcommand{\tev}{\ifthenelse{\boolean{inbibliography}}{\ensuremath{~T\kern -0.05em eV}\xspace}{\ensuremath{\mathrm{\,Te\kern -0.1em V}}}\xspace}
\newcommand{\gev}{\ensuremath{\mathrm{\,Ge\kern -0.1em V}}\xspace}
\newcommand{\mev}{\ensuremath{\mathrm{\,Me\kern -0.1em V}}\xspace}
\newcommand{\kev}{\ensuremath{\mathrm{\,ke\kern -0.1em V}}\xspace}
\newcommand{\ev}{\ensuremath{\mathrm{\,e\kern -0.1em V}}\xspace}
\newcommand{\gevc}{\ensuremath{{\mathrm{\,Ge\kern -0.1em V\!/}c}}\xspace}
\newcommand{\mevc}{\ensuremath{{\mathrm{\,Me\kern -0.1em V\!/}c}}\xspace}
\newcommand{\gevcc}{\ensuremath{{\mathrm{\,Ge\kern -0.1em V\!/}c^2}}\xspace}
\newcommand{\gevgevcccc}{\ensuremath{{\mathrm{\,Ge\kern -0.1em V^2\!/}c^4}}\xspace}
\newcommand{\mevcc}{\ensuremath{{\mathrm{\,Me\kern -0.1em V\!/}c^2}}\xspace}

\def\gsim{{~\raise.15em\hbox{$>$}\kern-.85em
          \lower.35em\hbox{$\sim$}~}\xspace}
\def\lsim{{~\raise.15em\hbox{$<$}\kern-.85em
          \lower.35em\hbox{$\sim$}~}\xspace}

\def\tell1  {TELL1\xspace}
\def\ukl1   {UKL1\xspace}

%% file: ourdefs.tex
\newcommand{\DsK}      {\ensuremath{\Ds\hspace{-0.0625em} K}\xspace}

\newcommand{\strong}	{\ensuremath{\delta}\xspace}

\newcommand{\rdsk}	{\ensuremath{r_{\DsK}}\xspace}

\ifthenelse{\boolean{uselhcbcpconvention}}%
{%
}{%

}%

%% file: introduction.tex
\section{Introduction}
Neutral beauty mesons may oscillate into their antiparticles, so that the 
physical states (those with well-defined masses and lifetimes) are admixtures of the flavor eigenstates. This mixing is parametrized by the magnitudes 
of the dispersive and absorptive components of the $\langle B | H |\bar{B}\rangle$ transition amplitude -- a box loop in the standard model (SM) -- as well as their relative phase, denoted $\phi_{d,s}$ in the $B^0$ and $B^0_s$ systems, respectively. The dispersive component generates mass splittings of the physical states, $\Delta m_{(d,s)}$, and is sensitive to heavy off-shell contributions from new physics (NP). The absorptive component generates width splittings $\Delta\Gamma_{(d,s)}$. It arises only from on-shell internal charm and up quark contributions, and is therefore less sensitive to possible NP effects.

%Neutral beauty mesons may oscillate into their antiparticles via loop level processes, so that the 
%physical states (those with well defined masses and lifetimes) are admixtures of the flavor eigenstates. This gives
%rise to both mass ($\Delta m_{(d,s)}$) and width ($\Delta\Gamma_{(d,s)}$) splittings in the $B^0_d$ and $B^0_s$ systems, as well
%as two types of related $CP$ violation. 

Two types of mixing-related $CP$ violation arise in these systems. The first, $CP$ violation in mixing, occurs when the meson and anti-meson have different
probabilities to oscillate into each other, and is predicted to be very close to zero in the SM with a very
small theoretical uncertainty. The second, $CP$ violation in the interference of decay and mixing, occurs when both the meson
and anti-meson can decay to the same final state, and the decay paths with and without intermediate mixing interfere. This kind of $CP$
violation is highly sensitive to the phase, $\phi_{d,s}$, and its absolute size depends on the final state in question.

Aside from their intrinsically fundamental nature, measurements of mass and width splittings and mixing-related $CP$ violation are
of great interest because many of the observables can be predicted very precisely in the SM, and because new particles or force-carriers
beyond the SM (BSM) can alter these predictions in experimentally observable ways. Many different experimental collaborations have contributed
to our understanding of mixing in the $B$ system, from the initial discovery of $B$ mixing by the ARGUS collaboration~\cite{Prentice:1987ap}, to precise measurements
of $\Delta m_{(d,s)}$ and the CKM-angles $\alpha$ and $\beta$ at the $B$-factories and Tevatron, to recent precise measurements
of $\phi_s$, $\Delta\Gamma_s$, and first precise mixing-related measurements of the CKM-angle $\gamma$ at ATLAS, CMS, and LHCb. At the same
time, great progress has been made in making more precise theoretical predictions of the SM values of many of these quantities, making
these experimental measurements sensitive probes of BSM physics.

The remainder of this document covers the current status of experimental measurements for each observable of
interest, as well as their near-term outlook. Section~\ref{sec:phisdgs} covers measurements of $\phi_s$ and $\Delta\Gamma_s$, section~\ref{sec:dmdgd} 
measurements of $\Delta m_{(s,d)}$. Section~\ref{sec:photpol} covers the measurements of photon polarization in radiative decays, while
sections~\ref{sec:alpha}~to~\ref{sec:gamma} cover measurements of the CKM angles $\alpha$, $\beta$, and $\gamma$, respectively.
For historical reasons, measurements of $CP$ violation in neutral meson mixing are covered in the proceedings of WG2~\cite{WG2PROC}. We then discuss
ongoing theoretical developments relevant to our WG in Sec.~\ref{sec:theory}. Finally we conclude,
and discuss the medium to long-term outlook for mixing-related measurements in the $B$ system.

%% file: phis-deltags.tex
\section{Measurements of $\phi_s$ and $\Delta\Gamma_s$}
\label{sec:phisdgs}

The $\phi_s$ weak phase and $B^0_s$ meson decay width difference can
be extracted by a time dependent full angular decay analysis of the 
$B^0_s \to J/\psi K^+ K^-$ decays. Such measurements have traditionally
used the subset of decays where the $K^+ K^-$ pair originates in the decay
of a $\phi$ meson, because its narrow width allows for an excellent background
rejection, but since the conference the first measurement which uses
the full $K^+ K^-$ mass spectrum~\cite{LHCbKKHighMass} has become available, and
more such measurements are expected in the future. The importance of
$\phi_s$ is that, in the absence of significant loop-diagram contributions
to $B^0_s \to J/\psi K^+ K^-$ decays (``penguin pollution''),
its value is precisely predicted~\cite{PHISSM} in the SM
\begin{equation}  
\phi_s = -0.036 \pm 0.002 \; ,
\end{equation}
and accurate experimental measurements of $\phi_s$ are therefore 
sensitive probes of the presence of BSM effects in the mixing and decay of $B^0_s$ mesons.

\begin{figure}
  \begin{center}
    \begin{tabular}{c c}
      \includegraphics[height=5.5cm]{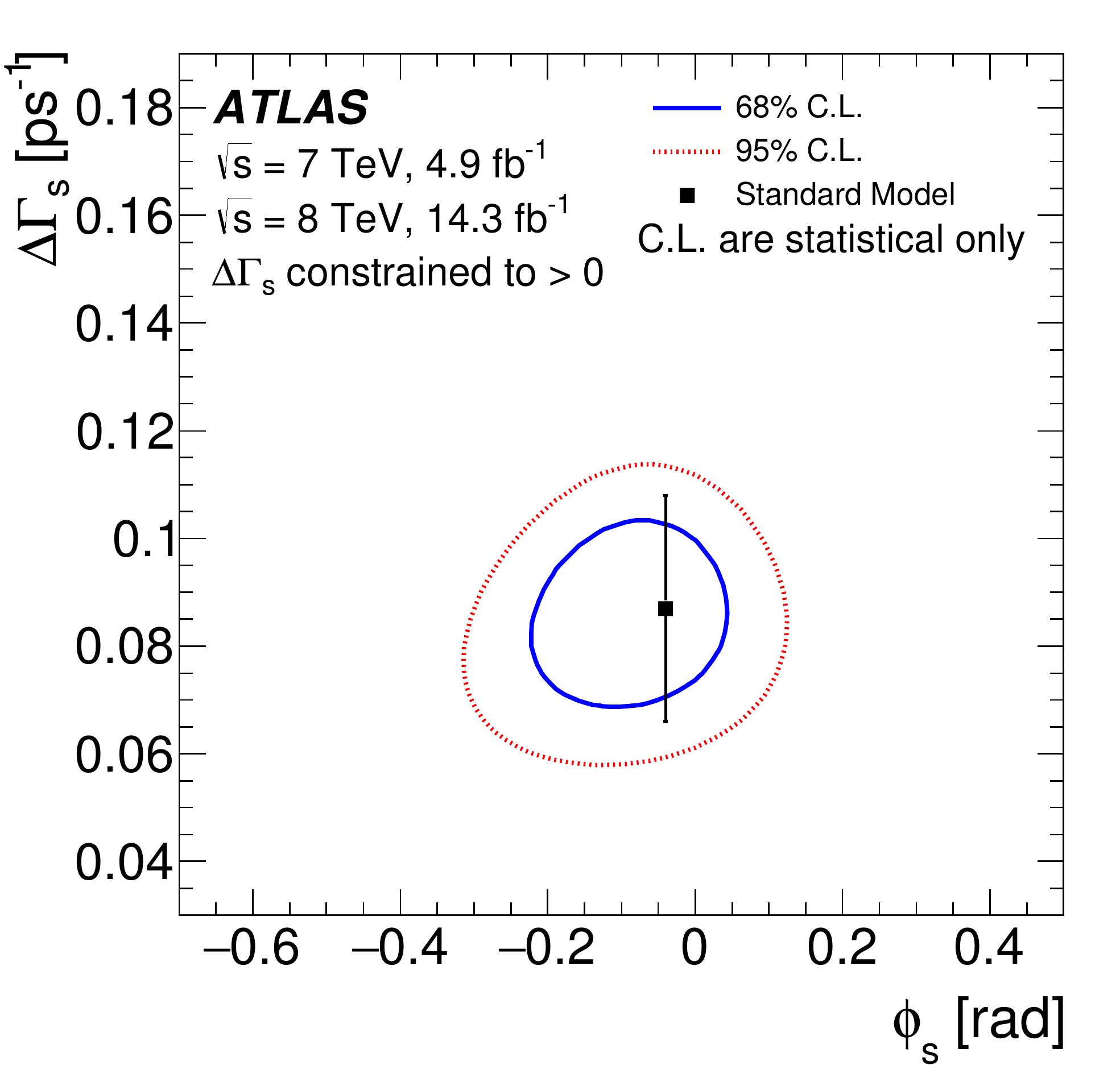} &
      \includegraphics[height=5.5cm]{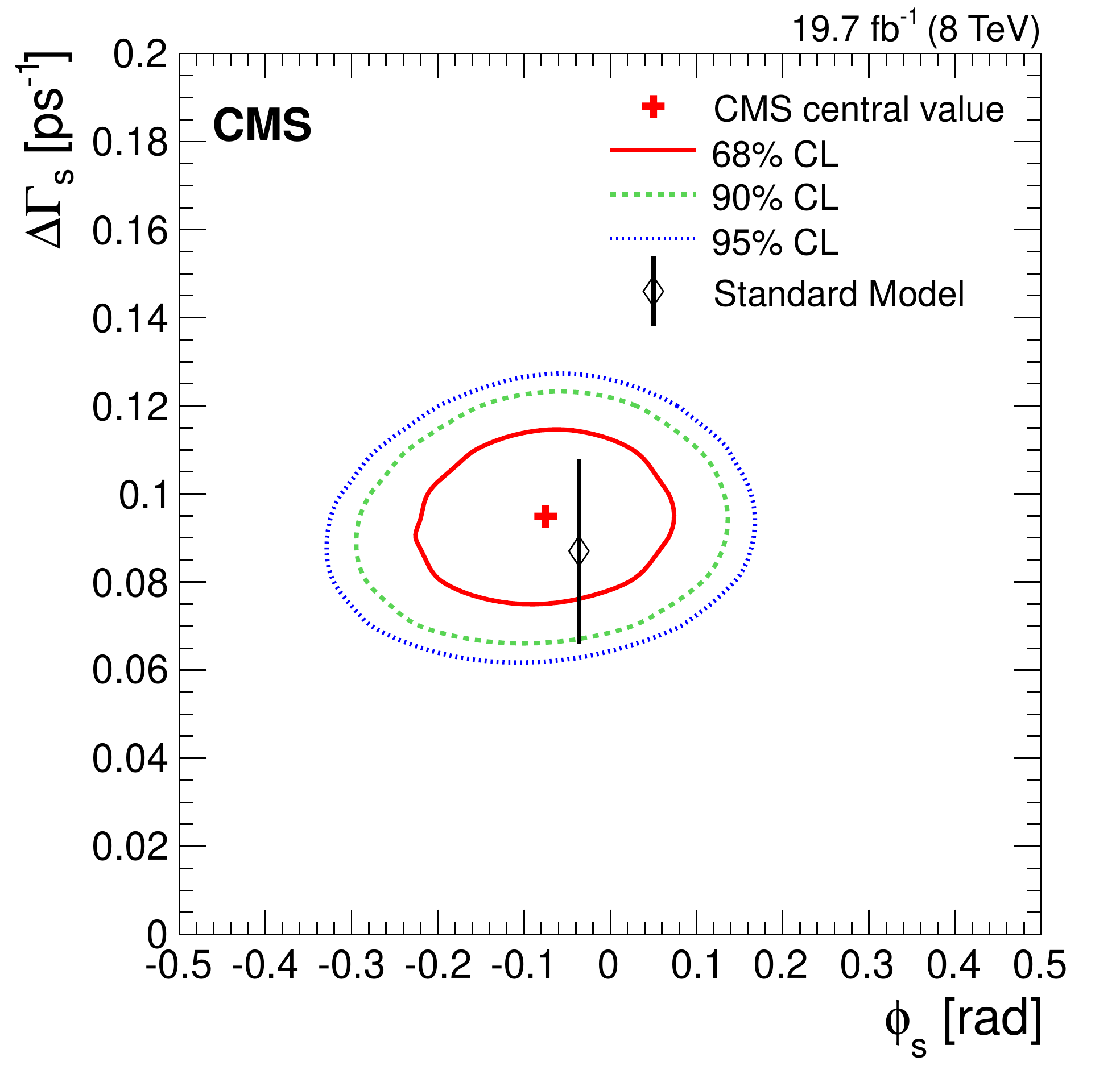} 
    \end{tabular}
  \end{center}
  \vspace{-0.5cm}
  \caption{\label{phis_atlas_cms}Results on $\phi_s$ and $\Delta\Gamma_s$ from
  $B^0_s \to J/\psi \phi$ decays from (left) ATLAS~\cite{atlas_phis_8TeV} and (right) CMS~\cite{cms_phis}, reproduced from the respective citations.}
\end{figure}

The results that the ATLAS~\cite{ATLASPHIS} and CMS Experiments obtained on their Run1
datasets using $B^0_s \to J/\psi \phi$ decays are:
\begin{eqnarray}
  \phi_s & = & -0.090 \pm 0.078 (\mbox{stat}) \pm 0.041 (\mbox{syst})\:[\mbox{rad}] \; , \\
  \Delta\Gamma_s & = & \phantom{-}0.085 \pm 0.011 (\mbox{stat}) \pm 0.007 (\mbox{syst})\:[\mbox{ps}^{-1}]  \; ,
\end{eqnarray}
for ATLAS \cite{atlas_phis_8TeV} and:
\begin{eqnarray}
  \phi_s & = & -0.075 \pm 0.097 (\mbox{stat}) \pm 0.031 (\mbox{syst})\:[\mbox{rad}]  \; , \\
  \Delta\Gamma_s & = & \phantom{-}0.095 \pm 0.013 (\mbox{stat}) \pm 0.007 (\mbox{syst})\:[\mbox{ps}^{-1}]  \; ,
\end{eqnarray}
for CMS \cite{cms_phis}. As can be seen in Fig.~\ref{phis_atlas_cms} both measurements are in
excellent agreement with the SM expectations.

The LHCb measurements of $\phi_s$ are covered in detail elsewhere in these proceedings~\cite{LHCBPHISPROC}.
LHCb uses not only $B^0_s \to J/\psi \phi$ but also $\Bs\to \jpsi\pi^+\pi^-$ decays,
where the $\pi^+\pi^-$ system is dominated by the $f_0(980)$ resonance and is greater than $97.7\%$ \CP-odd.
In this case no angular analysis is required, considerably simplifying the measurement
and providing important complementarity of systematic uncertainties.
The analysis of $B^0_s \to J/\psi \phi$ finds $\Delta\Gamma_s = 0.0805  \pm 0.0091         \pm  0.0032$~ps$^{-1}$,
where the first uncertainty is statistical and the second systematic. 
A combination of $B^0_s \to J/\psi \phi$ and $\Bs\to \jpsi\pi^+\pi^-$ gives $\phis = -0.010  \pm  0.039$~rad, which is the most 
precise single-experiment determination of this quantity.
LHCb has also measured $\phi_s$ using $\Bs\to \psi(2S)\phi$ decays, obtaining
$0.23^{+0.29}_{-0.28} \pm 0.02$ rad.

While $B^0_s \to J/\psi K^+ K^-$ is expected to be dominated by the tree-level diagram,
as the experimental precision on $\phi_s$ improves it will become increasingly important
to account for residual penguin pollution~\cite{PENGUINPOLTHEORY} in order to correctly interpret any agreement,
or otherwise, with the theoretical prediction. The size of such effects can be controlled
using a combination of $B^0_s \to J/\psi K^{*0}$ and $B^0 \to J/\psi \rho$ decays,
which are related by U-spin symmetry and in the limit of zero non-factorizable SU(3)
breaking are sensitive to the same, universal, penguin amplitudes and phases.
LHCb has performed measurements of both modes~\cite{LHCb-PAPER-2015-034,LHCb-PAPER-2014-058} and
measures essentially zero penguin pollution to $\phi_s$ with a precision of around 15~mrad
in all three polarization-dependent phases. The LHCb measurement is statistics limited
and penguin pollution to $\phi_s$ should therefore
remain under control as the experimental precision approaches the SM value.

In addition to comparisons with the SM prediction, such tree-dominated measurements of $\phi_s$ can also be compared with the measurement
of the related quantity $\phi_s^{ss\overline{s}}$ in the loop-dominated decays $\Bs \to \phi\phi$ and $\Bs\to \Kp\pi^-\Kp\pi^-$.
In the absence of BSM physics effects, this quantity is expected to be very close to zero.
LHCb has measured~\cite{LHCb-PAPER-2014-026} $\phi_s^{ss\overline{s}} = -0.17 \pm 0.15 \pm 0.03$ rad using
$\Bs\to \phi\phi$, in excellent agreement with the SM prediction. The measurement,
which uses $\Bs\to \Kp\pi^-\Kp\pi^-$, is ongoing; it is considerably more challenging and
requires a two-dimensional Dalitz analysis to account
for the interfering intermediate resonances. 

LHCb has also measured~\cite{LHCb:2017ood} the time-dependent $CP$ violation in $B^0 \to \pi^+ \pi^-$ and $B^0_s \to K^+ K^-$ decays.
These can be interpreted\footnote{In the absence of information about $B^0_s \to K^+ K^-$, measurements of time-dependent $CP$ violation in
$B^0 \to \pi^+ \pi^-$ have traditionally~\cite{HFLAV} been combined with other two-body $B^0$ decays in an isospin analysis to obtain a measurement of $\alpha$.}
in a combined U-spin analysis~\cite{Fleischer:1999pa} as either measurements of $\gamma$ or $\phi_s$.
Because the combined U-spin analysis is much less sensitive
to U-spin breaking~\cite{LHCb-PAPER-2014-045} when interpreted as a measurement of $\phi_s$, this is therefore the currently preferred way
to interpret the measurements of these $CP$ observables.

\begin{figure}
  \begin{center}
    \begin{tabular}{c c}
      \includegraphics[height=7.5cm]{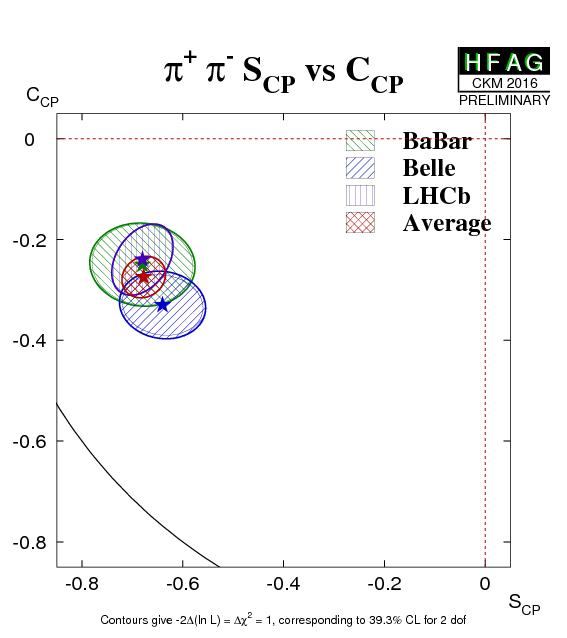} &
    \end{tabular}
  \end{center}
  \vspace{-0.5cm}
  \caption{\label{b2pipiwahfag}World average of time-dependent $CP$ observables in $B^0 \to \pi^+ \pi^-$, reproduced from HFLAV~\cite{HFLAV}.}
\end{figure}

$CP$ violation in $B^0 \to \pi^+ \pi^-$ has been measured previously by BaBar~\cite{Lees:2012mma} and Belle~\cite{Adachi:2013mae},
while the measurement of $B^0_s \to K^+ K^-$ is unique to LHCb. Using a two-dimensional fit to the mass and decay-time
of the neutral $B$ mesons, LHCb observes $CP$ violation in the interference of mixing and decay of $B^0_s$ mesons for the first time, 
and finds
\begin{eqnarray}
C_{\pi\pi}      &=& -0.24 \pm 0.07 \pm 0.01,\phantom{space}
S_{\pi\pi}      = -0.68 \pm 0.06 \pm 0.01,\phantom{space}
\\
C_{KK}      &=& \phantom{+}0.24 \pm 0.06 \pm 0.02,\phantom{space}
S_{KK}      = \phantom{+}0.22 \pm 0.06 \pm 0.02,\phantom{space}
\\
A^{\Delta}_{KK}      &=& -0.75 \pm 0.07 \pm 0.11,\phantom{space}
\end{eqnarray}
where the first uncertainty is statistical and the second systematic. It can be seen that the (co)sinusoidal $C$ and $S$ parameters have
negligible systematic uncertainties, while the hyperbolic $A^{\Delta}_{KK}$, which is sensitive to differences in the distributions
of $B^0_s$ and $\bar{B^0_s}$ induced by $\Delta\Gamma_s$ at high decay-times, is much more sensitive to experimental effects.
This is the most precise single-experiment
determination of $S_{\pi\pi}$, and significantly improves the world-average of this parameter as shown in Fig.~\ref{b2pipiwahfag}.
A combined interpretation in terms of $\phi_s$ is not available at the time of writing but is expected to be performed
in the future, and both LHCb and Belle-II are expected to improve our knowledge of the $CP$ observables in the future. 
The measurement of $A^{\Delta}_{KK}$ is a world first, allowing for a twofold reduction in the ambiguity of the $\phi_s$ determination
and can also be interpreted as a measurement of $\Delta\Gamma_s$.

%% file: deltam-deltagd.tex
\section{Measurements of $\Delta m_{(d,s)}$ and $\Delta\Gamma_d$}
\label{sec:dmdgd}

The neutral $B$ meson mass splittings $\Delta m_{(d,s)}$ have been precisely measured by BaBar, Belle, CDF, D0, and LHCb.
The world average~\cite{HFLAV} of these measurements is currently dominated by the LHCb determinations,
based on an analysis~\cite{LHCb-PAPER-2015-031} of semileptonic $B^0$ decays in the case of $\Delta m_d$, and based on an analysis~\cite{LHCb-PAPER-2013-006} of $B^0_s \to \D^-_s \pi^+$ decays
in the case of $\Delta m_s$. LHCb finds
\begin{equation}
\Delta m_d = (0.505 \pm 0.0021 \pm 0.0010) \textrm{ps}^{-1}\,,\phantom{space} 
\Delta m_s = (17.768 \pm 0.023 \pm 0.006) \textrm{ps}^{-1}\,,
\end{equation}
where the first uncertainty is statistical, and the second systematic. Neither analysis is systematics limited,
and LHCb is expected to update both measurements in the future. Belle-II will also be able to make a significant contribution
to a precise measurement of $\Delta m_d$. A measurement of $\Delta m_s$ will be difficult for ATLAS and CMS because of a lack of
efficient triggers for purely hadronic decays, but may become possible~\cite{PallaHLLHC} in the HL-LHC era once their first-level tracking triggers come online,
which would provide an important independent cross-check of LHCb's measurement.

The $B^0$ width splitting $\Delta\Gamma_d$ is predicted~\cite{Lenz:2011ti} to be $\Delta\Gamma_d/\Gamma_d = (4 \pm 1)\cdot 10^{-3}$ in the SM, and measuring it
is therefore a useful null test of the SM~\cite{Gershon:2010wx}. Such tests are particularly important as a non-null value of $\Delta\Gamma_d/\Gamma_d$ could have important
implications for the interpretation of $CP$ violation in the mixing of $B^0$ mesons, particularly in the context~\cite{Borissov:2013wwa} of the D0 dimuon asymmetry measurement.
LHCb has analyzed the effective decay-times of $B^0 \to J/\psi K^{*0}$
and $B^0 \to J/\psi K^0_\textrm{S}$ decays and obtains~\cite{LHCb-PAPER-2013-065} $\Delta\Gamma_d/\Gamma_d = (-4.4 \pm 2.5 \pm 1.1)\cdot 10^{-2}$,
while the current WA is dominated by the ATLAS measurement~\cite{Aaboud:2016bro}
of $\Delta\Gamma_d/\Gamma_d = (-0.1 \pm 1.1 \pm 0.9)\cdot 10^{-2}$, where the first uncertainties are statistical and the second systematic.
While the systematic uncertainties are dominated by simulation sample sizes, both measurements are still an order
of magnitude away from probing the SM prediction, so it will be important that even the subdominant
systematics scale with luminosity if we hope to one day observe a non-null $\Delta\Gamma_d/\Gamma_d$ at its SM value. 
 
The BaBar Collaboration studies the $B_d^0-\overline{B_d}^0$ oscillations to
test the conservation of the $CPT$ symmetry~\cite{babar_cpt}. At the lowest order in $|q/p|-1$
and $z$, the two mass eigenstates can be written:
\begin{eqnarray}
B_H & = & (p\sqrt{1+z} \; B^0 - q\sqrt{1-z} \; \overline{B}^0) / \sqrt{2} \,, \\
B_L & = & (p\sqrt{1-z} \; B^0 + q\sqrt{1+z} \; \overline{B}^0) / \sqrt{2} \,,
\end{eqnarray}
where:
\begin{equation}
  |q/p| = 1 - \frac{2 \Im(m_{12}^* \Gamma_{12})}{4|m_{12}|^2 + |\Gamma_{12}|^2}, \hspace{1cm}
  z = \frac{(m_{11}-m_{22}) - i (\Gamma_{11} - \Gamma_{22})/2}{\Delta m - i \Delta \Gamma /2}.
\end{equation}

The test is performed by fitting the $C$ and $S$ parameters of the $CP$ violation in the
interference between mixing and decay in $B \to c\bar{c} K^0_{S,L}$, when the other $B$
in the event decays to the $\ell^{\pm}X$ final state, separating the cases when the decay
of the $B$ to the $CP$-eigenstate happens before or after the decay of the other $B$ to
the flavor-specific final state. The assumption $|\overline{A}/A|=1$, where $A$ ($\overline{A}$)
is the amplitude of $B^0 \to c\bar{c} K^0$ ($\overline{B}^0 \to c\bar{c} \overline{K}^0$) is
not enforced.

The final result is:
\begin{eqnarray}
  \Im (z) & = &\phantom{-} 0.010 \pm 0.030 (\mbox{stat}) \pm 0.013 (\mbox{syst}) \, ,\\
  \Re (z) & = & -0.065 \pm 0.028 (\mbox{stat}) \pm 0.014 (\mbox{syst}) \, ,\\
  |\overline{A}/{A}| & = & \phantom{-} 0.999 \pm 0.023 (\mbox{stat}) \pm 0.017 (\mbox{syst}) \, ,
\end{eqnarray}
in good agreement with $CPT$ conservation.

%% file: photpol.tex
\section{Measurements of photon polarization in radiative decays}
\label{sec:photpol}
Radiative decays such as $\B^0_s \to \phi \gamma$ give access to information
about the photon polarization, and tests the presence of possible right-handed
couplings and set constraints on the Wilson coefficients $C_7$ and $C_7^{\prime}$ in EFT frameworks.
Previous measurements~\cite{HFLAV} of photon polarization in radiative $B^0$ decays
have been compatible with SM predictions. 

\begin{figure}
  \begin{center}
    \begin{tabular}{c c}
      \includegraphics[height=5.5cm]{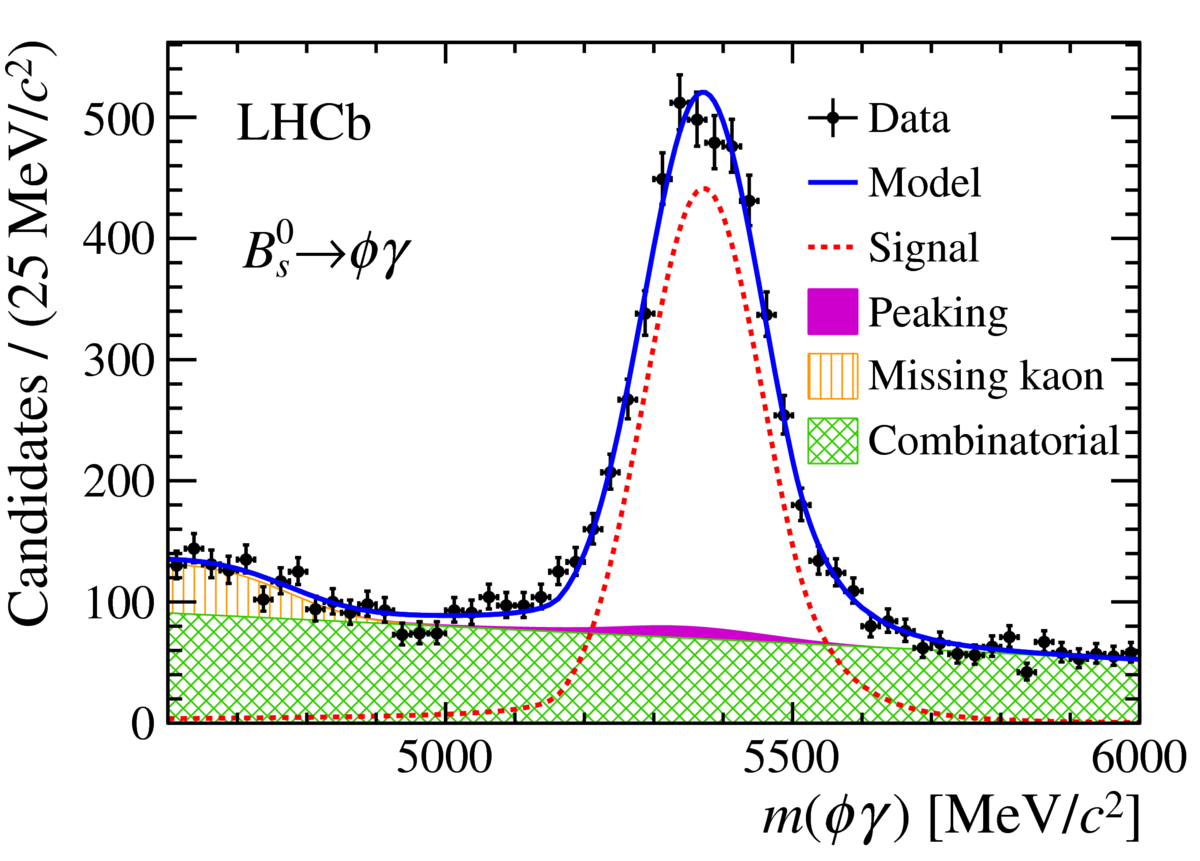} &
      \includegraphics[height=5.5cm]{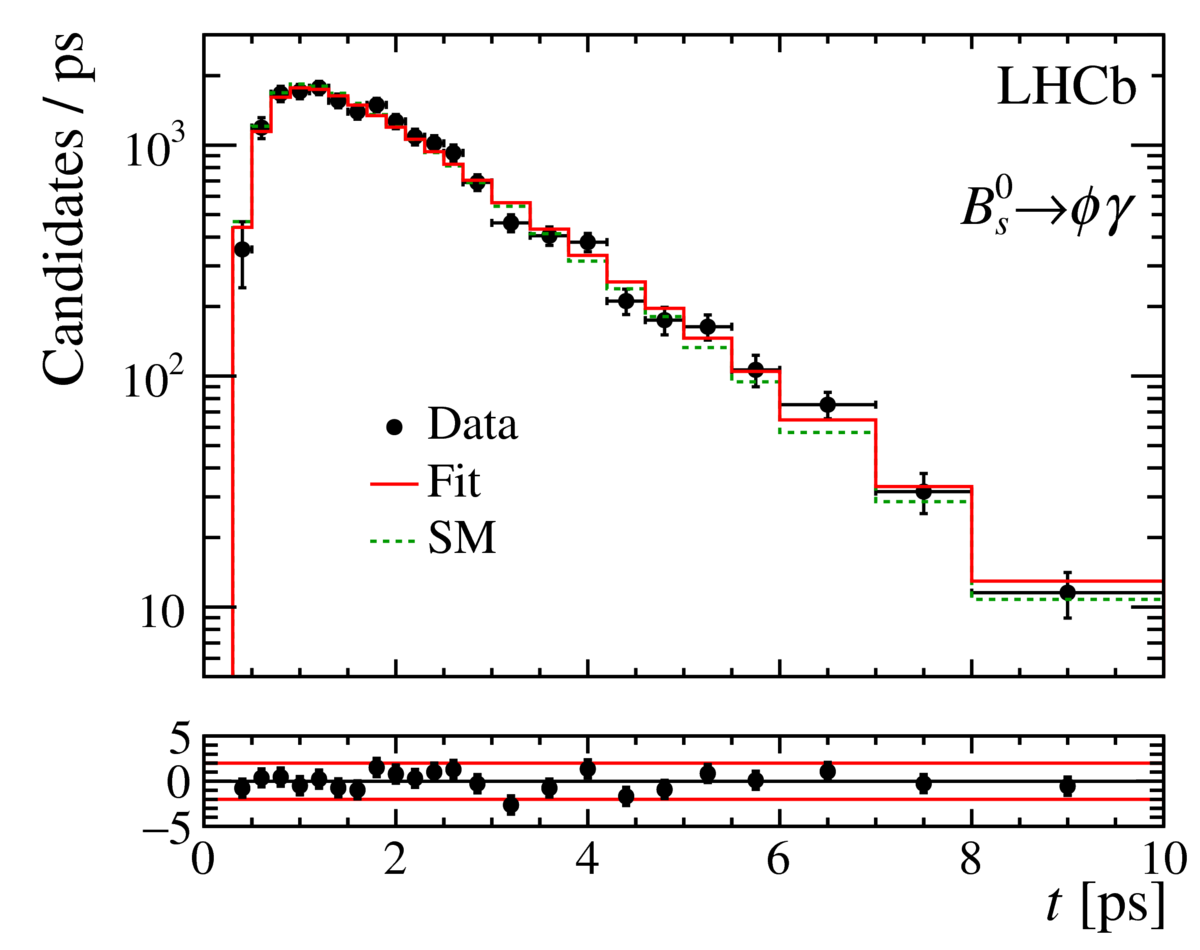} 
    \end{tabular}
  \end{center}
  \vspace{-0.75cm}
  \caption{\label{phigammamasstime}Distributions of the fitted $\B^0_s \to \phi \gamma$ (left) invariant mass and (right) decay-time, reproduced from~\cite{LHCb-PAPER-2016-034}.}
\end{figure}

LHCb is uniquely able to probe photon polarization in radiative $B^0_s$ decays,
which offer a complementary test of BSM physics. While a full analysis requires
tagging the production flavor of the $B^0_s$ meson, the large value of $\Delta\Gamma_s$
allows a measurement of the effective lifetime of $\B^0_s \to \phi \gamma$ to be
interpreted in terms of the $CP$-violating observable $A^\Delta$. LHCb
performs this measurement~\cite{LHCb-PAPER-2016-034} by normalizing the $\B^0_s \to \phi \gamma$ decay-time distribution
to that of $\B^0 \to K^{*0} \gamma$, cancelling most experimental biases on the decay-time.
The fitted mass and decay-time distributions are shown in Fig.~\ref{phigammamasstime}; LHCb finds
$A^\Delta = -0.98^{+0.46}_{-0.52}$, in $2\sigma$ agreement with the SM
prediction~\cite{Muheim:2008vu} of $A^\Delta = 0.047 \pm 0.025 \pm 0.015$. The measurement is statistics
limited, and future tagged analyses will enable the (co)sinusoidal $CP$ observables to
be measured, allowing both the real and imaginary components of $C_7$ and $C_7^{\prime}$ to be constrained.

%% file: alpha.tex
\section{Measurements of the CKM angle $\alpha$}
\label{sec:alpha}

One of the possible ways to extract the CKM angle $\alpha$ is through
an isospin analysis of the $B \to \rho\rho$ decays. The Belle Collaboration
recently updated their measurements of the branching fraction, longitudinal
polarization fraction, and $CP$-violation parameters of the $B^0 \to \rho^+\rho^-$
decay, using their full dataset \cite{belle_rhorho}. The results are:
\begin{eqnarray}
  \mathcal{B} (B^0 \to \rho^+\rho^-) & = & [28.3 \pm 1.5 (\mbox{stat}) \pm 1.5 (\mbox{syst})] \times 10^{-6} \; , \\ 
  f_L & = & 0.988 \pm 0.012 (\mbox{stat}) \pm 0.023 (\mbox{syst}) \; , \\
  S & = & -0.13 \pm 0.15 (\mbox{stat}) \pm 0.05 (\mbox{syst}) \; , \\
  C & = & \phantom{-}0.00 \pm 0.10 (\mbox{stat}) \pm 0.06 (\mbox{syst}) \; .
\end{eqnarray}
Combining the results summarized above with previous $B \to \rho\rho$
measurements in an isospin analysis, one of the solutions is:
\begin{equation}
  \alpha = (93.7 \pm 10.6)^{\circ} \; ,
\end{equation}
in very good agreement with the SM.

%% file: beta.tex
\section{Measurements of the CKM angle $\beta$}
\label{sec:beta}

\begin{figure}
  \begin{center}
    \begin{tabular}{c c}
      \includegraphics[height=7.5cm]{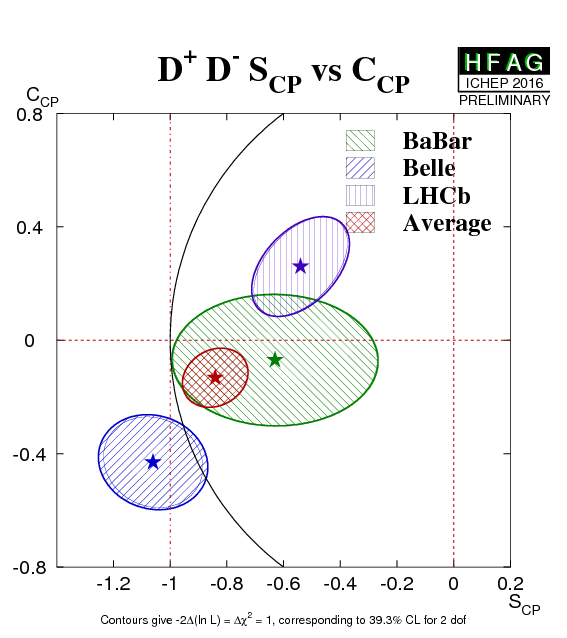} &
    \end{tabular}
  \end{center}
  \vspace{-0.5cm}
  \caption{\label{b2ddwa}World average of time-dependent $CP$ observables in $B^0 \to D^+ D^-$, reproduced from HFLAV~\cite{HFLAV}.}
\end{figure}

While the world average~\cite{HFLAV} of $\beta$ is still dominated
by the BaBar~\cite{Aubert:2009aw} and Belle~\cite{Adachi:2012et} measurements, LHCb also contributes~\cite{LHCb-PAPER-2012-035} and can be expected
to reach the BaBar and Belle individual sensitivities once Run~II data and
improvements in flavor tagging are included in the analysis. LHCb has also measured~\cite{LHCb-PAPER-2015-005}
$CP$ violation in $B^0_s \to J/\psi K^0_S$ decays,
which can help to constrain the size of penguin contributions to the measurement
of $\beta$ from $B^0 \to J/\psi K^0_S$, as well as the $CP$ violating parameters in
$B^0 \to D^+ D^-$ decays, which can also be interpreted in terms of constraints on $\beta$.
The current world average of measurements of $B^0 \to D^+ D^-$ is shown in Fig.~\ref{b2ddwa}; 
interestingly while all measurements are compatible with each other, the LHCb measurement
does not confirm the maximal (within uncertainty) Belle measurement of $S$.

The BaBar and Belle Collaboration obtained the first observation of $CP$-violation
in the channels $B^0 \to D^{(*)}_{CP} h^0$, where $h^0$ is a light unflavored neutral
meson, by combining their final datasets \cite{babar_belle_D0h0}. Neglecting the
very small contribution from the CKM suppressed amplitude $b \to u\bar{c}d$,
the time dependent analysis is sensitive to $\sin(2\beta)$. A joint likelihood,
sharing the same physics parameters but independent background modeling parameters,
the two experiments find:
\begin{eqnarray}
-\eta_f S & = & +0.66 \pm 0.10 (\mbox{stat}) \pm 0.16 (\mbox{syst}), \\
C & = & +0.02 \pm 0.07 (\mbox{stat}) \pm 0.03 (\mbox{syst}), 
\end{eqnarray}
with the significance of the observation being 5.4 standard deviations. 

%% file: gamma.tex
\section{Measurements of the CKM angle $\gamma$}
\label{sec:gamma}
Measurements of the CKM angle $\gamma$ require the interference of $b\to u$ and $b\to c$ transitions.
Such interference can occur in the decays of charged as well as neutral $B$ hadrons, in tree-dominated
transitions as well as transitions where both tree and loop diagrams contribute.
The measurement of $\gamma$ from the tree-dominated decays of $B^\pm$ mesons, covered in the 
proceedings of WG5~\cite{WG5PROC}, has a particular importance as it allows a purely tree-level\footnote{Higher order box corrections~\cite{ZupanBrodGamma}
only enter at $\delta\gamma/\gamma \approx 10^{-7}$, well beyond any current or future experimental sensitivity.}
determination of the apex of the unitarity triangle, and therefore a test of the self-consistency of the CKM
mechanism of $CP$ violation when compared with determinations of $\alpha$ and $\beta$ in transitions where
both tree and loop diagrams contribute. It is also, however, possible to measure $\gamma$ by exploiting
$CP$ violation in the interference of mixing and decay of neutral $B$ mesons. These time-dependent determinations
of $\gamma$ are particularly powerful in the case of $B^0_s$ mesons, because the large width difference between
the light and heavy $B^0_s$ eigenstates makes additional $CP$ observables available compared
to the $B^0$ system and allows for a determination of $\gamma$ with fewer ambiguities. 
%While these are not
%strictly tree-level determinations, as the mixing of neutral $B$ mesons

A particularly powerful time-dependent measurement of $\gamma$, described in detail in these proceedings~\cite{DSKPROC}, utilizes the decay $B^0_s \to D^\pm_s K^\mp$.
In this case the $b\to u$ and $b\to c$ transitions are both of order $\lambda^3$, leading to large interference, and
the large value of $\Delta\Gamma_s$ allows for a determination of $\gamma$ with only a twofold ambiguity. 
The preliminary result obtained with the full Run~I LHCb dataset, which shows $3.6\sigma$ evidence for $CP$-violation in this mode, is
\begin{equation}
\gamma      = (127_{-22}^{+17})^\circ\,,\phantom{space}
\strong = (  358_{-16}^{+15})^\circ\,,\phantom{space}
\rdsk   = 0.37_{-0.09}^{+0.10}\,,\phantom{space}
(68.3\% \textrm{CL})\phantom{space}  
\end{equation}
\begin{equation}
\gamma      = (127_{-50}^{+33})^\circ\,,\phantom{space}
\strong = (  358_{-33}^{+31})^\circ\,,\phantom{space}
\rdsk   = 0.37_{-0.19}^{+0.19}\,,\phantom{space}
(95.4\% \textrm{CL})\phantom{space}  
\end{equation}
where $\strong$ is the $CP$-conserving angle between the $b\to u$ and $b\to c$ transitions,
$\rdsk$ is the amplitude ratio of the interfering diagrams, and the intervals for the angles are expressed modulo $180^\circ$.
The uncertainties are a combination of statistical and systematic ones; the statistical uncertainties dominate
and all systematic uncertainties are expected to scale with luminosity for the foreseeable future.

While not the most sensitive single-mode determination of $\gamma$, $B^0_s \to D^\pm_s K^\mp$ plays a similar
role in the overall LHCb combination~\cite{LHCb-PAPER-2016-032} of $\gamma$ to that of the GGSZ measurement. Because of their twofold ambiguity, these measurements
select the ``correct'' solution among the ones allowed by the most precise ADS/GLW measurement~\cite{LHCb-PAPER-2016-003}. For this reason the determination
of $\gamma$ from $B^0_s \to D^\pm_s K^\mp$, which is only possible at LHCb, will remain a key measurement for both the current
and upgraded LHCb detectors. LHCb is also pursuing a measurement of time-dependent $CP$-violation in
the decay mode $B^0 \to D^\pm \pi^\mp$, described in these proceedings~\cite{BDPIPROC}, but no results are available yet.
This measurement, previously performed by BaBar~\cite{Aubert:2005yf}, \cite{Aubert:2006tw} and Belle~\cite{Bahinipati:2011yq}, \cite{Ronga:2006hv}
is much less sensitive than $B^0_s \to D^\pm_s K^\mp$, both because of smaller interference and because
the small value of $\Delta\Gamma_d$ leads to fewer accessible $CP$-observables. The much smaller $CP$ asymmetry
also makes this measurement particularly sensitive the asymmetries in the flavor tagging of $B^0$ and $\bar{B^0}$ mesons.
Nevertheless, it is expected that both LHCb and Belle-II will carry out this measurement in the future.

%% file: theory.tex
%% Please use the skeleton file you have received in the
%% invitation-to-submit email, where your data are already
%% filled in. Otherwise please make sure you insert your
%% data according to the instructions in PoSauthmanual.pdf
%\documentclass{PoS}
%
%\usepackage{amssymb, amsmath}
%
%\title{Summary of Working Group 4 : mixing and mixing-related $CP$ violation in the $B$ system}
%
%\ShortTitle{All SM everything}
%
%\author{\speaker{Alessandro Gaz}\\
%        Nagoya\\
%        E-mail: \email{gaz@hepl.phys.nagoya-u.ac.jp}}
%\author{
%        \speaker{Vladimir V. Gligorov}\thanks{I would like to thank my mum, dad, and trade union representative for their support. Couldn't have done it without you!}\\
%        LPNHE, Universit\'{e} Pierre et Marie Curie, Universit\'{e} Paris Diderot, CNRS/IN2P3, Paris, France\\
%        E-mail: \email{vgligoro@lpnhe.in2p3.fr}
%        }
%\author{
%        \speaker{Dean Robinson}\thanks{Show me an ambulance dashing off into the night, and I shall show you the theorists bravely in pursuit.}\\
%        University of Cincinnati, Cincinnati, Ohio, USA\\
%        E-mail: \email{dean.robinson@uc.edu}\\
%        }
%
%%\author{Another Author\\
%%        Affiliation\\
%%        E-mail: \email{...}}
%
%\abstract{$B$ mesons mix and in mixing do wonderful and amazing things.}
%
%\FullConference{9th International Workshop on the CKM Unitarity Triangle\\
%		28  November - 3 December 2016\\
%		Tata Institute for Fundamental Research (TIFR), Mumbai, India}
%
%
%\begin{document}
\section{Theory Developments}
\label{sec:theory}

%for $\Delta M_{d,s}$, and therefrom $V_{ts}$ and $V_{td}$.

\subsection{Lattice Calculations}

The most precise computation to date for $|V_{td}|$ and $|V_{ts}|$, or alternatively $\Delta m_{d,s}$, is recently available from the Fermilab/MILC collaboration~\cite{Bazavov:2016nty}, building on previous results for $B$ mixing matrix elements~\cite{Aoki:2014nga,Carrasco:2013zta,Bazavov:2012zs,Albertus:2010nm,Gamiz:2009ku,Dalgic:2006gp}. These predictions are achieved via a three flavor lattice QCD  calculation for the neutral $B$ mixing hadronic matrix elements, $\langle \bar{B} | \mathcal{O}^q_i | B\rangle$, with $\mathcal{O}^q_i$ combinations of scalar, pseudoscalar, vector and axial vector four-quark operators as required by the theory of interest. In the SM, the oscillation frequency
\begin{equation}
	\label{eqn:SMDM}
	\Delta m_q = \frac{G_F^2 m_W^2 M_{B_q}}{6 \pi^2} S_0(m_t^2/m_W^2) \eta_{2 B}|V_{tb}V^*_{tq}| f_{B_q}^2 \hat{B}^{(1)}_{B_q}\,,
\end{equation}
in which $\hat{B}^{(1)}_{B_q}$ is a renormalization-improved bag parameter associated with the operator product of left-handed quark currents $\mathcal{O}^q_1 = \bar{b}^\alpha\gamma^\mu(1-\gamma^5) q^\alpha~\bar{b}^\beta \gamma^\mu(1-\gamma^5) q^\beta$, while $S_0(x_t)$ and $\eta_{2B}$ encode known electroweak and perturbative QCD corrections, respectively. Computation of $f_{B_q}\sqrt{\hat{B}^{(1)}_{B_q}}$ may be combined with either direct measurements of $\Delta m_{s,d}$ or CKM global fits to test the self-consistency of data with lattice calculations. One may also extract $|V_{td}/V_{ts}|$ or $\Delta m_d/\Delta m_s$ via the flavor $SU(3)$ breaking ratio 
\begin{equation}
\xi = \sqrt{f_{B_s}^2 \hat{B}^{(1)}_{B_s}/ f_{B_d}^2 \hat{B}^{(1)}_{B_d}}\,,
\end{equation}
in which many theory uncertainties cancel. Computation of the other $B$ mixing matrix elements, corresponding to the other operators $\mathcal{O}^q_i$, permit similar predictions for BSM theories.

This lattice calculation is performed for $N_f = 2+1$ light quark flavors. Non-perturbative renormalization effects are included, while two-loop chiral-continuum extrapolation of the lattice results determines the physical limits. Presentation of all crucial details may be found in Ref.~\cite{Bazavov:2016nty}. Present results are
\begin{gather}
	f_{B_d}\sqrt{\hat{B}^{(1)}_{B_d}} = 227.7(9.5)(2.3)~\text{MeV}\,, \qquad f_{B_s}\sqrt{\hat{B}^{(1)}_{B_s}} = 274.6(8.4)(2.7)~\text{MeV}\,,\nonumber\\
		\xi = 1.206(18)(6)\,,
\end{gather}
currently the most precise predictions to date. Combination of these results with $|V_{tq}|$ results from CKM global fits yields 
\begin{gather}
	\Delta m^{\text{CKM}}_d = 0.630(53)(42)(5)(13)~\text{ps}^{-1}\,, \qquad \Delta m^{\text{CKM}}_s = 19.6(1.2)(1.0)(0.2)(0.4)~\text{ps}^{-1}\,,\nonumber \\
	\Delta m^{\text{CKM}}_d/\Delta m^{\text{CKM}}_s = 0.0321(10)(15)(0)(3)\,,
\end{gather}
in an approximately $2\sigma$ tension with direct measurements of these parameters \cite{HFLAV}
\begin{equation}
	\Delta m^{\text{HFLAV}}_d = 0.5064(19)~\text{ps}^{-1}\,,\qquad \Delta m^{\text{HFLAV}}_s = 17.757(21)~\text{ps}^{-1}\,.
\end{equation}
Alternatively, using the direct measurements for $\Delta m_{d,s}$ produces CKM predictions
\begin{gather}
	|V_{td}| = 8.00(33)(2)(3)(8)\times 10^{-3}\,,\qquad |V_{ts}| = 39.0(1.2)(0.0)(0.2)(0.4)\times 10^{-3}\,,\nonumber\\
	|V_{td}/V_{ts}| = 0.2052(31)(4)(0)(10)\,,
\end{gather}
approximately $\sim2\sigma$ below the results from global CKM fits~\cite{Charles:2004jd, Charles:2015gya}. In particular, in the context of the $b \to d$ unitarity triangle, this tension manifests as a tension between the allowed regions for the CP violating parameter $\epsilon_K$ and $\Delta m_{d}/\Delta m_s$, with potentially interesting theory implications. 

These same lattice calculations also generate predictions for $B_{d,s} \to \mu\mu$ and $\Delta\Gamma_{d,s}$, with some mild tensions for $B_d$ decays. Further improved calculations for $N_f = 2 + 1 + 1$ that include the charm quark sea effects and physical quark masses are anticipated.

\subsection{Constrained MFV theories}
Given the possible tensions in the unitarity triangle between $\epsilon_K$ and $\Delta m_{d,s}$, it is informative to consider which classes of BSM theories could account for such tension. A class involving near-minimal BSM contributions are models of constrained minimal flavor violation (CMFV) \cite{Buras:2000dm,Buras:2003jf,Blanke:2006ig,Blanke:2006yh,Blanke:2016bhf}. In these theories, the SM Yukawa couplings $Y_u$ and $Y_d$ are treated as $U(3)\times U(3)$ flavor violating spurions, generating the sole source of $CP$ violation.  BSM effects are encoded in higher dimensional SM effective operators.

This class of theories preserves the unitarity structure of the CKM matrix. The precisely measured CKM matrix elements for the first two generations imply the relation, at percent level precision,
\begin{equation}
	R_t \equiv \bigg|\frac{V_{tb}^*V_{td}}{V_{cb}^*V_{cd}}\bigg| \simeq \frac{|V_{td}/V_{ts}|}{\lambda}\,,
\end{equation}
where $\lambda$ is the usual Wolfenstein parameter. The ratio $|V_{td}/V_{ts}|$ is determined precisely via Eq.~\eqref{eqn:SMDM} from direct measurements of $\Delta m_d/ \Delta m_s$ and lattice computations of $\xi$. Thus, measurements of $\Delta m_{d,s}$ and the time-dependent $CPV$ observable $S_{\psi K_S}$, which determines $\sin(2\beta)$ (see below), fully determine a `universal unitarity triangle' (UUT) for the $b \to d$ system for all CMFV theories. 

The electroweak loop function $S_0(x_t)$ in $\Delta m_{d,s}$~\eqref{eqn:SMDM} also appears in the CP violating parameter $\epsilon_K$. CMFV replaces $S_0(x_t)$ with a generalized universal function $S(v)$, bounded below by $S_0(x_t)$ in most compelling BSM scenarios. The UUT is, however, independent of the electroweak loop function, as it drops out of the $\Delta m_{d}/\Delta m_s$ ratio. Consequently, in CMFV theories, there is an extra degree of freedom between the predictions for $|V_{ub}/V_{cb}|$ and the CKM angle $\gamma$, both fixed by the UUT, and the constraints from measurements of $\epsilon_K$. 

This extra freedom is, however, found to be insufficient to relax the tension between $\Delta m_{d,s}$ and $\epsilon_K$ generated in the FNAL/MILC lattice results, when applied together with the bound $S(v) \ge S_0(x_t)$ \cite{Blanke:2016bhf, Blanke:2016xvd}. One may take either $\Delta m_{s,d}$ or $\epsilon_K$ direct measurements as inputs, and thereby determine all other CKM matrix elements as functions of $S(v)$ via the UUT constraints: $\Delta m_{d,s}$ direct measurements imply an upper bound on $\epsilon_K$ that is too small compared to data; $\epsilon_K$ data implies lower bounds on $\Delta m_{d,s}$, that are above current measurements. If the tension between lattice and experimental data persists in the unitarity triangle, it will become imperative to consider new sources of flavor violation in $\Delta F = 2$ processes, beyond CMFV models.

\subsection{Precision control of Penguin Pollution}
The time-dependent $CPV$ observable $S_{\psi K_S}$, generated by the interference of $B$-mixing and $B_d \to J/\psi K_S$ decay amplitudes, has long been considered a golden mode for extraction of $\phi_d$, via the relation $S_{\psi K_S} = \sin (\phi_d + \delta \phi_d)$. The latter `penguin pollution' phase is expected to be CKM and loop-suppressed, yielding a clean measurement of $\phi_d \simeq 2 \beta$ in the SM. Similar techniques may be used to extract $\phi_s$ from, e.g., $B_s \to J/\psi \phi$. Estimates place $\delta \phi_d \lesssim 1^\circ$ \cite{Boos:2004xp,Ciuchini:2005mg,Li:2006vq,Faller:2008zc,Ciuchini:2011kd,Jung:2012mp,DeBruyn:2014oga,Frings:2015eva,Ligeti:2015yma}, which is near to or larger than the expected precision of upcoming measurements, requiring theoretical control of these terms.

In general terms, two different paths have been followed to control these penguin pollution terms. Direct calculational approaches have been attempted using QCD factorization~\cite{Boos:2004xp}, or factorization combined with perturbative QCD~\cite{Li:2006vq}, that yield very small estimates $\delta \phi_d \lesssim 0.1^\circ$. A more recent study~\cite{Frings:2015eva} makes use of an OPE-type approach to directly calculate the penguin pollution contributions. This analysis integrates out the $u$-quark loop, associated with this penguin pollution term, on the basis that the typical momentum flow is large. One obtains a factorization formula for the penguin contributions, and one may show that soft and collinear divergences formally cancel or factorize up to $\Lambda_{\text{qcd}}/m_{J/\psi}$ corrections. Large-$N_c$ arguments permit estimation of remaining uncertainties, yielding $\delta \phi_d < 0.68^\circ$ and $\delta \phi_s \lesssim 1^\circ$. 

A second, complementary path makes use of light quark flavor symmetries to constrain or eliminate penguin pollution effects in particular observables. For example on may make use of $U$-spin symmetry to  control pollution of $S_{\psi K_S}$ with $B \to J/\psi \pi$ data, up to symmetry breaking effects~\cite{Ciuchini:2005mg,Faller:2008zc,Ciuchini:2011kd,DeBruyn:2014oga}. Full $SU(3)$ analyses \cite{Jung:2012mp,DeBruyn:2014oga, Ligeti:2015yma} that treat all flavor symmetry breaking effects, or subsets of them, can be used to control penguin pollution. A simultaneous fit~\cite{Jung:2012mp} predicts $\delta \phi_d \lesssim 0.6^\circ$.  Flavor symmetry relations beyond the flavor symmetry limit~\cite{Ligeti:2015yma} permit extraction of $\phi_d$ with penguin pollution suppressed by $SU(3)$ breaking, but require precision measurement of $CP$-averaged rates in the charged and neutral $B \to J/\psi K$ and $B \to J/\psi \pi$ systems.

\subsection{Lessons of isospin violation}
Measurement of CP averaged rates at the percent-level precision required for the use of above-mentioned flavor $SU(3)$ results \cite{Jung:2012mp, Ligeti:2015yma} in turn requires careful inclusion of isospin violating effects in the ratio of neutral and charged $B$ production at the $\Upsilon(4S)$~\cite{Jung:2015yma}. Knowledge of this ratio, $r_{+0} = f_{+-}/f_{00}$, is crucial for precision measurement of branching ratios:  the measured charged (neutral) $B$ branching ratios are known up to a $1 + r_{+0}$ ($1 + 1/r_{+0}$) factor, expected to be of the order of a few percent from unity. Control of this ratio can be achieved, independently of isospin assumptions, via e.g. counting single versus double tagged semileptonic $B$ decays~\cite{Aubert:2005bq}, from which $r_{+0}$ may be extracted under the assumption $f_{00} + f_{+-} = 1$ and uncorrelated reconstruction efficiencies, or via e.g. measuring relative branching ratios for inclusive semileptonic processes~\cite{Hastings:2002ff,HFLAV}, in which isospin violation is expected to be suppressed to order $1/m_{b}^2$. Combination of available data for these two approaches yields $r_{+0} = 1.027 \pm 0.037$~\cite{Jung:2015yma}.

%\end{document}

%% file: conclusion.tex
\section{Conclusion}
The last two decades have seen enormous progress in the understanding of $B$ meson mixing and mixing-related
$CP$ violation, both in terms of precise experimental measurements of the underlying constants of nature and in terms of 
the theoretical understanding of their values within the Standard Model. We now have precise measurements 
or stringent limits on the mass splitting, width splitting, and mixing phase in both the $B^0$ and $B^0_s$ systems,
while mixing-induced $CP$ violation is being precisely measured or constrained in an ever increasing number of final states.
With the LHCb upgrade~\cite{Bediaga:1443882} and Belle~II~\cite{Abe:2010gxa} detectors due to come online in the next few years,
and none of the fundamental measurements
yet systematically limited, we can expect this progress to continue. Important contributions, particularly as regards
$\phi_s$ and $\Delta\Gamma_d$ can also be expected from CMS and ATLAS, and in particular it is realistic to expect the $B^0_s$ mixing
phase to be measured significantly away from zero even at the Standard Model value within the next decade. Recently
the LHCb collaboration has proposed a Phase~II upgrade of its detector~\cite{Aaij:2244311}, to take data in the HL-LHC period,
which would collect 300~fb$^{-1}$, and enable not only a single-experiment observation of $\phi_s$ in multiple independent
decay modes, but also make it possible to see evidence for a non-zero $\Delta\Gamma_d$ at its Standard Model value.

Computation and further control of subleading corrections in the extraction of CKM angles and matrix elements from $B$ mixing has proceeded apace with new state-of-the-art lattice calculations. The former, particularly extraction of $\sin 2\beta$ from $b \to c\bar{c} s$ decays, has been the subject of various recent studies, with the emerging conclusion that theory uncertainties are likely smaller than the attainable experimental precision. Recent direct calculations as well as new flavor symmetry analyses will provide complementary handles to decide this question. State-of-the-art lattice results for $B$-mixing matrix elements indicate a possible evolving tension between predicted and measured mass splittings $\Delta m_{d,s}$, or in the context of the $b \to d$ CKM unitarity triangle, a tension between $\Delta m_{s,d}$ and $\epsilon_K$ allowed regions. Apart from constrained minimal violation theories, which seem unable to account for this tension, at present potential theoretical implications are relatively unexplored. 

%Particle physics is now roughly where the French monarchy was in 1788, and CERN is Versailles. But who will be our Robespierre?